# 14-m aperture deployable off-axis far-IR space telescope design for SALTUS observatory

**Daewook Kim,**[a,b] **Youngsik Kim,**[a] **Heejoo Choi,**[a] **Marcos Esparza,**[a] **Oliver Wu,**[a] **Yuzuru Takashima,**[a] **Art Palisoc,**[c] **Christopher Walker**[b,*]

[a]James C. Wyant College of Optical Sciences, University of Arizona, 1630 E. University Blvd., Tucson, AZ 85721
[b]Department of Astronomy and Steward Observatory, University of Arizona 933 N Cherry Ave., Tucson, AZ 85721
[c]L'Garde, Inc., 15181 Woodlawn Avenue, Tustin, CA 92780, USA

**Abstract**. The Single Aperture Large Telescope for Universe Studies (SALTUS) is a deployable space telescope designed to provide the astrophysics community with an extremely large far-infrared (far-IR) space observatory to explore our cosmic origins. The SALTUS observatory can observe thousands of faint astrophysical targets, including the first galaxies, protoplanetary disks in various evolutionary states, and a wide variety of solar system objects. The SALTUS design architecture utilizes radiatively cooled, 14-m diameter unobscured aperture, and cryogenic instruments to enable both high spectral and spatial resolution at unprecedented sensitivity over a wavelength range largely unavailable to any existing ground or space observatories. The unique SALTUS optical design, utilizing a large inflatable off-axis primary mirror, provides superb sensitivity, angular resolution, and imaging performance at far-IR wavelengths over a wide ±0.02° × 0.02° Field of View. SALTUS' design, with its highly compact form factor, allows it to be readily stowed in available launch fairings and subsequently deployed in orbit.

*Christopher Walker, E-mail: cwalker@arizona.edu

## 1 Introduction

The Single Aperture Large Telescope for Universe Studies (SALTUS) is a deployable far-infrared (far-IR) space observatory mission concept optimized to observe and investigate our cosmic origins. The space telescope shown in Fig. 1 operates from a halo orbit around L2. It will conduct groundbreaking studies by observing the first galaxies, protoplanetary disks at various evolutionary stages, and a wide variety of solar system objects, utilizing the superb photon collecting power of its unprecedented 14-m diameter unobscured clear aperture [1].

The telescope's unique optical design incorporates a radiatively cooled, inflatable 14-m primary mirror and cryogenic instruments. This design provides both high spectral and spatial resolution with high sensitivity across a wide wavelength range that is largely unexplored by existing ground or space observatories. These far-IR observation capability will bridge the knowledge gap between the local and distant universe using cryogenic coherent detectors (hot electron bolometer mixers—HEBs) and incoherent detectors (kinetic inductance detectors—KIDs), covering a wavelength from 34 to 659 µm [2].

The sunshield, depicted in Fig. 1, allows the 14-meter off-axis primary mirror to radiatively cool to below 45K. This feature is essential for the telescope's ability to capture far-infrared radiation with minimal background noise, further enhancing its wide spectrum observational coverage.

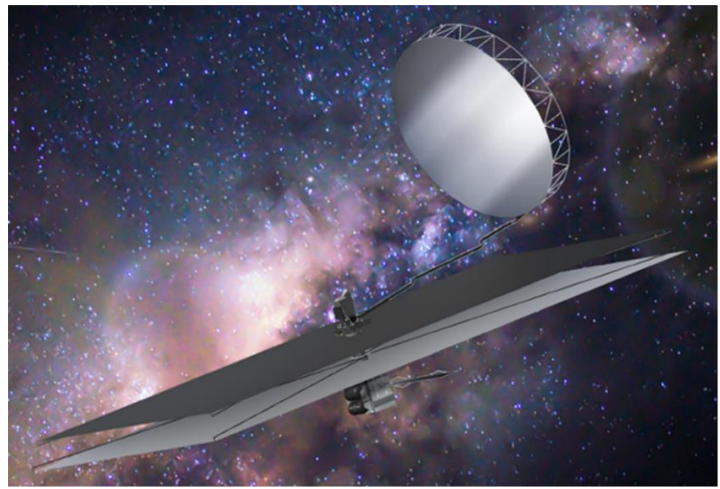

**Fig. 1** 3D rendered image of the SALTUS observatory, featuring an inflatable 14-m diameter off-axis primary mirror technology and its sunshield for radiative cooling below 45K.

SALTUS features an inflatable membrane [3 – 5] as its primary mirror (M1), which is connected to the spacecraft via a deployable single boom. The inflatable membrane is made from space-qualified polyimide film, designed to withstand critical strain under low pressures in a highly stable and predictable manner, as verified through finite element analysis and thermal vacuum (TVAC) testing.

The shape configurations of the deflated M1 are determined using an inverse problem solver, a method developed by L'Garde [5]. Given specific design inflation pressure and the membrane's material and geometric properties, the initial deflated profile is calculated to ensure the inflated M1 meets the optical design's target surface figure. The modeling process includes numerical integration of a first-order nonlinear differential equation with boundary conditions. The simulation of the initial deflated shape for the forward transformation is performed using Finite-element Analysis of Inflatable Membranes (FAIM), a geometric nonlinear membrane finite element simulation platform [5].

1-m and 3-m scale inflatable mirrors shown in Fig. 2 have been designed, modeled, and built to test the as-built system's performance and modeling fidelity. The 1-m diameter prototype was tested in a TVAC chamber at Northrop Grumman, as illustrated in Figure 2 (left). The mirror mount was designed to adjust its alignment and position relative to the deflectometry metrology system for the in-situ optical testing during the TVAC test [6].

The 1-m prototype was oriented with its mechanical axis parallel to the cylindrical axis inside the TVAC chamber. A custom-built inflation control unit maintained the internal pressure with a resolution of 10 Pa throughout the entire TVAC test. The internal pressure was set to 520 Pa, relative to the chamber pressure. Chamber pressure ranged from atmospheric (101,325 Pa) to near-vacuum (0.110 Pa) [6]. The deflectometry TVAC test results successfully confirmed the stable M1 surface shape and calibration capabilities. These metrology capabilities cover the low-order surface shape to the mid-spatial frequency surface errors. Furthermore, the 3-m scale model, illustrated in Fig. 2 (right), demonstrated the scalability of the inflatable mirror technology [2]. It highlighted the practical engineering perspective and the deterministic modeling approach for realizing a 14-m scale membrane mirror surface.

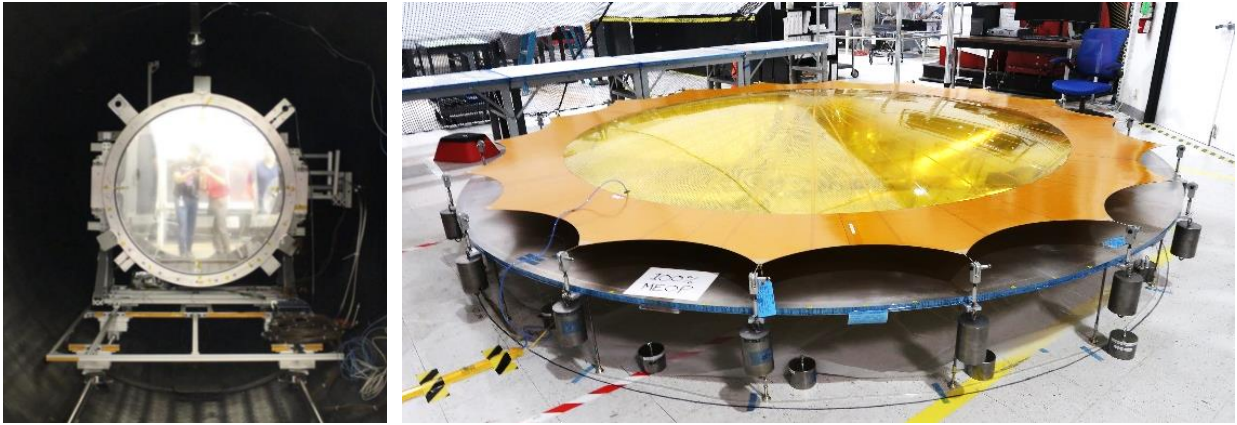

**Fig. 2** The 1-m scale SALTUS M1 prototype (left) in the TVAC testing chamber at Northrop Grumman [4] and the 3-m scale inflatable mirror (right) demonstrated at L'Garde.

## 2    Deployable Optical Design

*2.1 Deployable Optical Design Using Inflatable Primary Mirror*

The deployable SALTUS optical design, featuring a 14-m clear aperture off-axis primary mirror, is illustrated in Fig. 3. The inflatable M1, connected to the Cold Corrector Module (CCM) via a single boom structure, focuses the input beam into the CCM. This module is compactly designed to correct residual aberrations within a $\pm 0.02° \times 0.02°$ Field of View (FoV). The compactness of the CCM, in conjunction with the deployable boom and the inflatable M1, enables a small stowage form factor for SALTUS.

Unlike traditional optical design practices, where designers can specify the primary mirror as an asphere or a freeform surface type (e.g., using polynomials), the SALTUS M1 surface requires realizable and accurately modeled inflatable surface shapes. The SALTUS team addressed the inverse problem to model and optimize these shapes. This involved using the desired final surface shape from optical ray tracing outcomes, inflation pressure and control, and the material and geometric properties of the membrane to calculate the initial uninflated shape. We then employed the finite element software FAIM, validated against analytical solutions and experimental results [5, 6], to perform a shape accuracy analysis of the SALTUS M1. The final M1 shape resembles a near off-axis-parabolic surface and its ~38-mm PV (Peak-to-Valley) aspheric departure from best-fit sphere is depicted in Fig. 4. We have summarized the optimized optical prescription for each SALTUS mirror in Table 1.

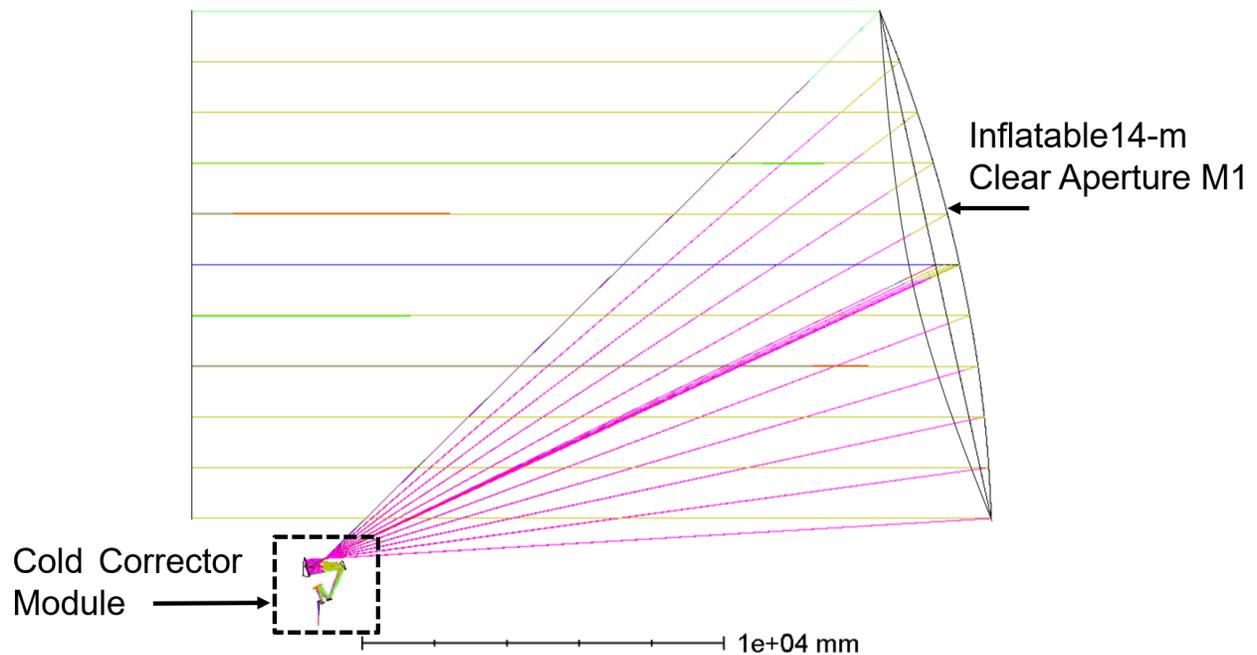

**Fig. 3** The SALTUS optical design features an unobscured, off-axis, 14-m clear aperture inflatable M1 and a highly compact Cold Corrector Module. This design enables a small stowage form factor for the SALTUS during launch.

The final modeled shape of M1 is integrated into the ZEMAX model to optimize and finalize the SALTUS optical design. This integration ensures that the downstream CCM can compensate for the as-inflated actual shape of M1. Consequently, the SALTUS optical design fully embodies the comprehensive knowledge and results of numerical modeling of the inflatable membrane mirror. Any remaining errors from modeling uncertainties and on-orbit environmental factors will be corrected by a low-order deformable mirror M3. The nominal optical design parameters for the diffraction-limited 14-m SALTUS telescope are outlined in Table 2.

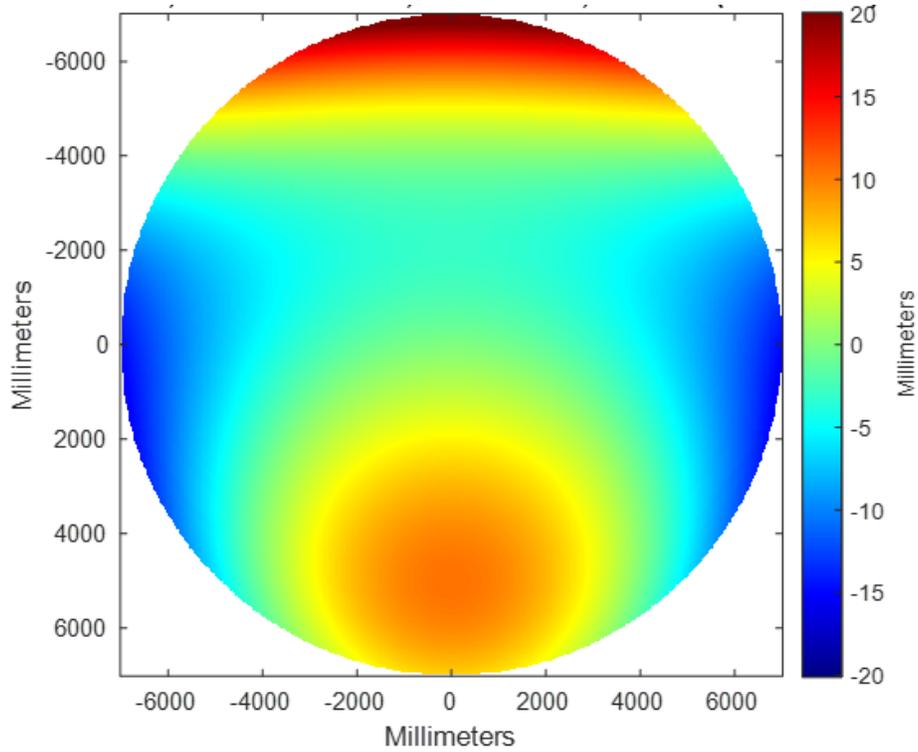

**Fig. 4** The aspheric departure map of the SALTUS M1 surface from the best-fit sphere is optimized using the inflatable surface finite element software, FAIM. The 37.978 mm Peak-to-Valley (PV) and 6.695 mm Root-Mean-Square (RMS) statistics illustrate its target design as an off-axis primary mirror with balancing astigmatism.

Table 1 SALTUS optical surface prescriptions

| Mirror No | Note | Radius of Curvature | Conic Constant k | Off-axis Distance | Clear Aperture | Surface Type |
|---|---|---|---|---|---|---|
| | | mm | - | mm | mm | - |
| M1 | Inflatable optic | 37,158.000 | -1 | 8,105.26 | 14,000 dia. | Inflated |
| M2 | | 927.640 | -1.00782 | -235.00 | 480 dia. | Aspheric |
| M3 | Deformable mirror | -316.313 | -4.98502 | 5.00 | 220 dia. | Aspheric |
| M4 | Folding flat | Infinite | 0 | N/A | 120 dia. | Flat |
| M5 | | -1,309.079 | 2.52388 | -615.00 | 220 × 120 | Aspheric |
| M6 | Field scanner | Infinite | 0 | N/A | 80 dia. | Flat |
| M7 | Fast steering mirror | Infinite | 0 | N/A | 80 dia. | Flat |

Table 2 Overall SALTUS optical design parameters

| Design Parameter | Nominal Value | Note |
| --- | --- | --- |
| Telescope Design Type | Unobscured off-axis design | Inflatable Telescope. |
| Field of View | $\pm 0.02° \times 0.02°$ | Scanned by M6 field scanner. |
| F/# | 16 | Working F/# matching the science instruments. |
| M1 Clear Aperture | 14-m diameter | Inflatable M1 membrane mirror. |
| M1 Surface Shape | Near off-axis parabola | FAIM-based surface modeling [5]. |
| Design Wavelength | 30 µm | Shortest science wavelength. All reflective achromatic. |
| Deformable Mirror | M3 | Compensates M1 shape error. Based on a matured legacy concept [9]. |
| Field Scanning Mirror | M6 | Flat Mirror. |
| Fast Steering Mirror | M7 | Flat Mirror. |
| Strehl Ratio | 0.82 - 0.94 | The center field Strehl ratio is 0.93. |

*2.2 Cold Corrector Module Design*

The Cold Corrector Module (CCM) shown in Fig. 5 is compactly packaged to balance the M1 aberration across the entire Field of View (FoV). Its design, necessitated by the challenging requirement to fit within the rocket's payload fairing volumes, adopts a highly off-axis folded configuration. This significantly deviates from traditional Three Mirror Anastigmat (TMA) designs, such as those used in the James Webb Space Telescope (JWST) [7] or the UArizona's 6.5-m monolithic space telescope [8], which are often employed for large FoV applications.

M2 creates an image conjugate plane (i.e., pupil plane) for M1 near the M3 location. M3 is strategically placed slightly off the exact pupil plane. This positioning aims to minimize the size of the convex M3 ensuring it still fits within the CCM envelope. Fabricating and testing convex mirrors can be challenging, but this approach offers the best technical value per cost. Furthermore, placing M3 exactly at the pupil plane would not fully correct the entire M1 as-inflated surface shape error and field common aberrations. This is due to the pupil aberration caused by the large off-axis and compact optical design of M2 and M3. The residual error is corrected by the aspheric, smaller M5 mirror downstream.

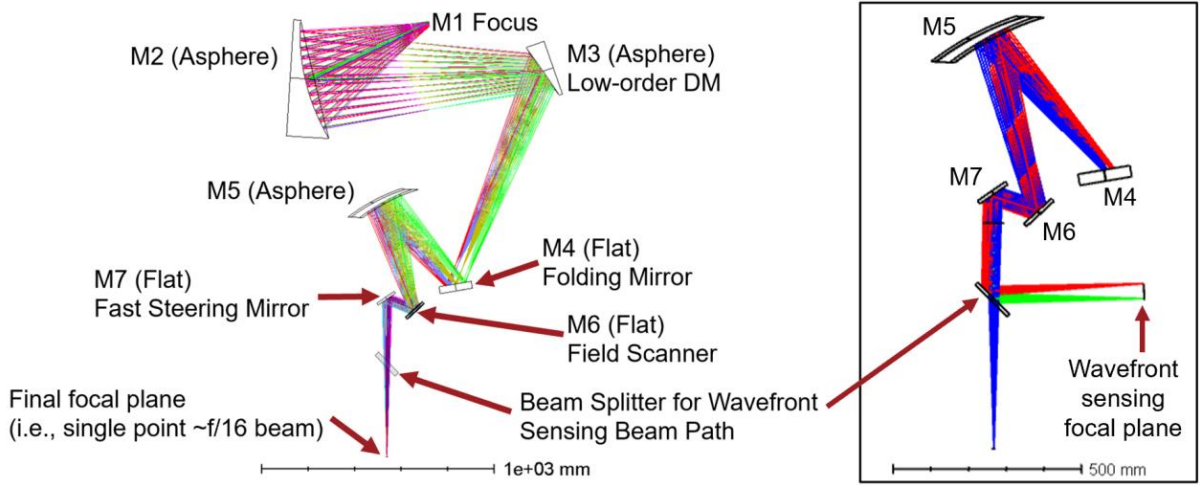

**Fig. 5** (Left) SALTUS's Cold Corrector Module, compact in design, corrects M1 aberration across the entire Field of View (FoV) and scans the FoV. (Right) The static beam splitter in the SPGD-based wavefront sensing path is depicted, showcasing the on-axis aligned science beam path in blue, the nominal monitoring 5 µm wavelength beam path in green, and the misaligned monitoring beam path example in red.

The SALTUS design achieves diffraction-limited nominal performance within a compact envelope by utilizing small mirrors in a folded geometry. It's noteworthy that these mirrors can be rapidly manufactured with standard meter-class Single Point Diamond Turning (SPDT) machines. These machines are highly deterministic and capable of producing freeform surfaces using matured slow or fast tool servo techniques, suitable for far-IR wavelength applications. Thus, the final shapes of M2 and M3 will be finely adjusted using an SPDT machine to compensate for the metrology data of the as-manufactured and as-inflated M1 shape. This nominal offset significantly reduces the dynamic range required of the M3 Deformable Mirror (DM) shown in Fig. 6.

The M3 is a mechanically simple quasi-static DM with a matured design legacy [9], avoiding complicated electronic or high-speed control loops. A 350-mm aperture DM that utilizes a stepper motor combined with differential ball screw actuators was successfully demonstrated and provided an excellent cross-coupling-free actuator response function with 0.025 µm surface positioning resolution and total 40 µm deflection range [9]. The simplicity of the mechanical design, demonstrated performance using as-built system, and relevant mirror size (i.e., 350-mm diameter which is larger than the 220-mm SALTUS M3) with sufficient control resolution (i.e., 0.025 µm which is much smaller than 30 µm the shortest science wavelength of SALTUS), significantly lowers the risk associated with deploying an adaptive optical system in orbit. While typical deformable mirrors correct low-to-mid spatial frequency errors, the SALTUS M3, using 19 linear actuators (stepper motors), primarily addresses low-spatial frequency errors, such as coma and astigmatism.

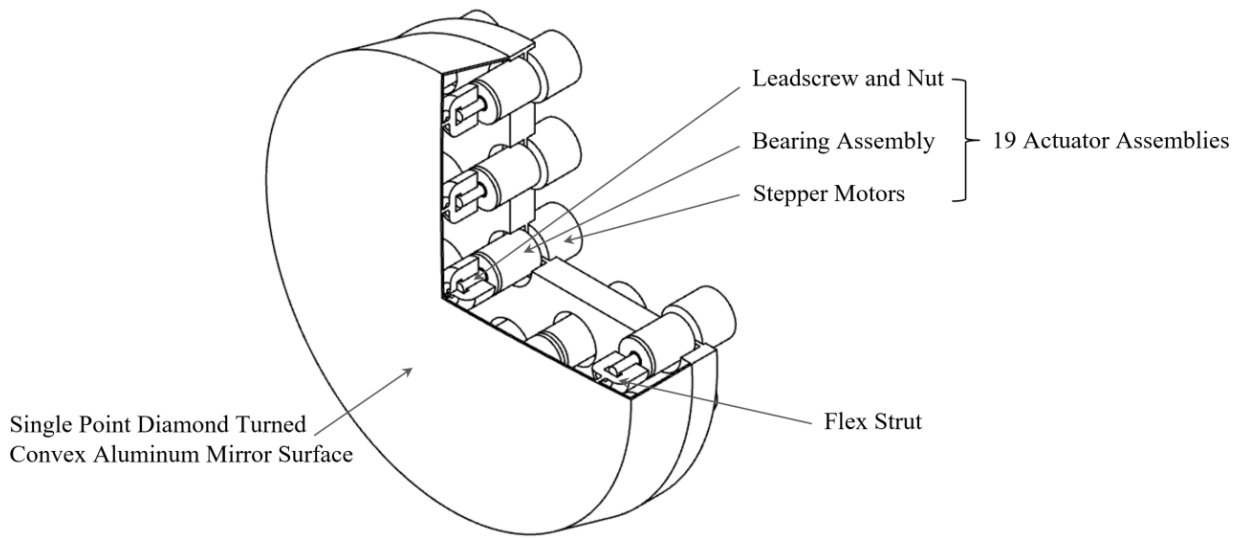

**Fig. 6** Schematic view of the 220-mm in diameter, 19-actuator, low-order SALTUS quasi-static deformable mirror (M3), directly benchmarking against the legacy design and hardware demonstration by Henderson and Gunn (funded by Department of Energy # ED-78-C-08-1551) [9].

The low-order DM enhances the resilience of the SALTUS design architecture. It compensates for modeling uncertainties, such as the shape differences between the on-ground and on-orbit M1 shapes, through its surface deformation. The M3 can also address other residual aberrations caused by misalignment and other optical surface deformations. Unlike common adaptive optics DMs, which operate at very high speeds (e.g., 1 – 2 kHz) during science observations for ground-based observatory applications, this DM is primarily utilized during the commissioning phase. Its main function is to adjust for the environmental impacts on the SALTUS observatory's optical performance. Optionally, M3 can also be employed during science observations to correct any slowly varying aberrations, such as those caused by thermal structural misalignment.

A standard model-free Stochastic Parallel Gradient Descent (SPGD) algorithm [10 – 13] evaluates the Root Mean Square (RMS) spot size across all observation fields on the monitoring detector planes shown in Fig. 5 (right). This evaluation is used to adjust the telescope's alignment and the low-order shape control of the M3 DM. The algorithm randomly adjusts the degrees of freedom (DoF) parameters in the telescope, with subsequent changes in spot size dictating the next set of DoF adjustments. Initial broad perturbations of the DoFs help quickly ascertain the alignment status during the early phases of alignment and collimation. A wavefront aberration model-based SPGD approach then refines the alignment and M3 shape precisely, enabling in-situ adjustments during scientific image acquisition if necessary.

Typically, the SPGD algorithm can reduce the wavefront control loop to less than $1\lambda_{SPGD}$ PV (Peak-to-Valley) [12, 13]. As depicted in Fig. 5 (right), ~5 µm wavelength light is used by the SPGD algorithm to sense wavefront errors based on the point spread functions. This corresponds to sensing wavefront errors down to $1/6\ \lambda_{Science}$ PV for the shortest 30 µm science wavelength, adequately driving the low-order M3 DM.

SALTUS's coherent and incoherent science instruments utilize feedhorn and lens coupled detectors. The M6 field scanner, located at the pupil conjugate plane, slowly scans the Field of View (FoV) to direct a stationary final science beam towards the science instruments' final focal plane [2]. The required M6 motion range is ± 2.5° in x-tilt and ± 2.9° in y-tilt. No translation motion is required when the pivot point is located at the center of the M6 mirror surface. The M7 Fast Steering Mirror corrects the high temporal frequency vibrations to stabilize the final Point Spread Function (PSF).

*2.3 Diffraction-Limited Nominal Optical Performance*

The spot diagrams in Fig. 7 depict the stationary final focal plane (i.e., the focal plane interfacing with the downstream science instruments) and show highly stable beam footprints across the entire Field of View (FoV), specifically ± 0.02° × 0.02°, achieved using the M6 field scanner located at the pupil plane.

Furthermore, this field scanner is instrumental in compensating and calibrating the on-orbit SALTUS observatory during the commissioning phase. The ability of the M6 field scanner to focus and feed all field point beams within the Airy Disk at the shortest design wavelength of 30 µm (illustrated by the black circle) demonstrates its capability to sequentially steer the beam toward the single feedhorn detector during science observations.

Figure 8 presents the PSF for the best and worst field points, showcasing the diffraction-limited nominal performance of the SALTUS design, which includes the inflatable primary M1 shape model and the CCM. The Strehl Ratios, ranging from ~ 0.82 to 0.94, indicate the high-quality imaging performance of the system. Notably, this diffraction-limited nominal imaging performance exceeds the requirements of the SALTUS spectroscopic science instruments (e.g., the GUSTO instrument [14]), providing a considerable margin for tolerancing analysis. This directly contributes to a robust space observatory architecture [2].

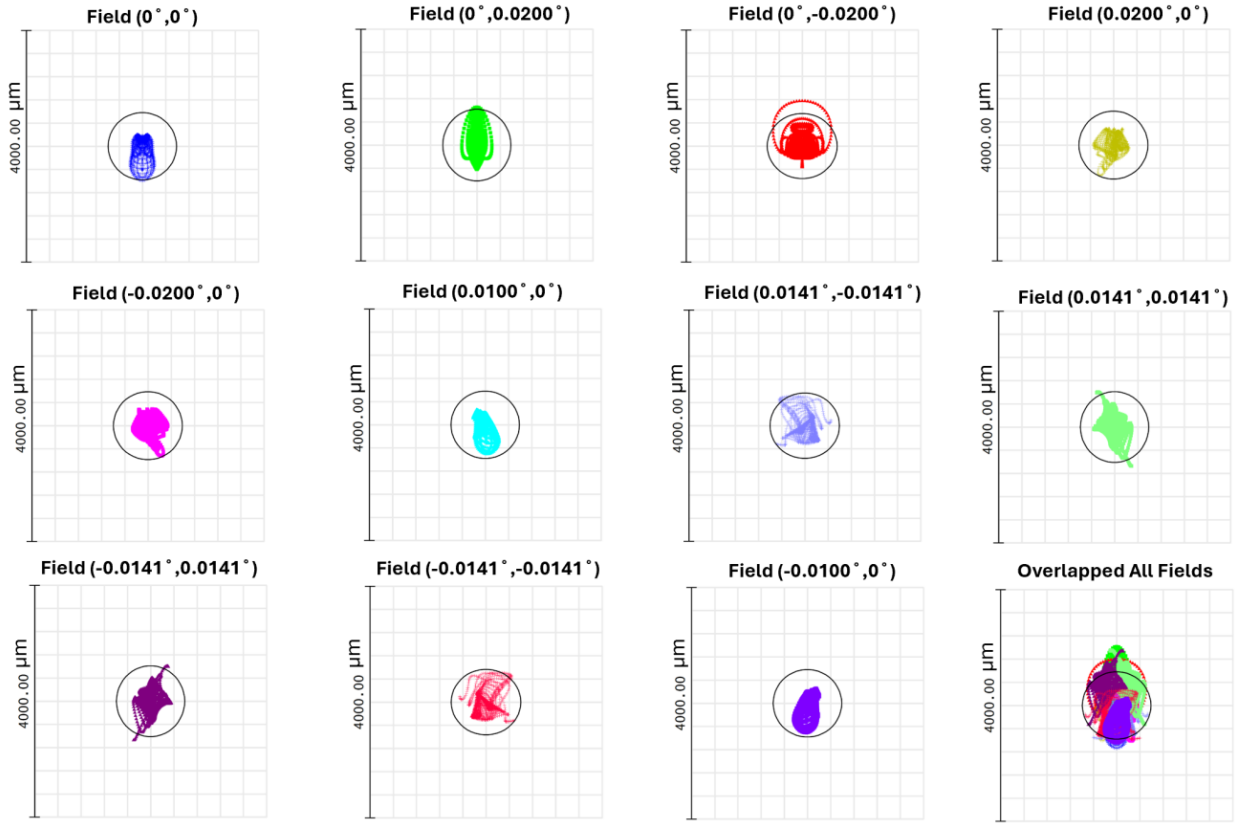

**Fig. 7** The F/16 beam spot diagrams, color-coded for all Field of View (FOV) points, illustrate how the beam is fed to the downstream far-infrared (far-IR) instruments using various single feedhorn detectors. The bottom-right corner of the diagram displays the overlapped stationary footprints of all spots using the Field Scanning Mirror M6. The black circle represents the Airy disk of 1.17 mm diameter.

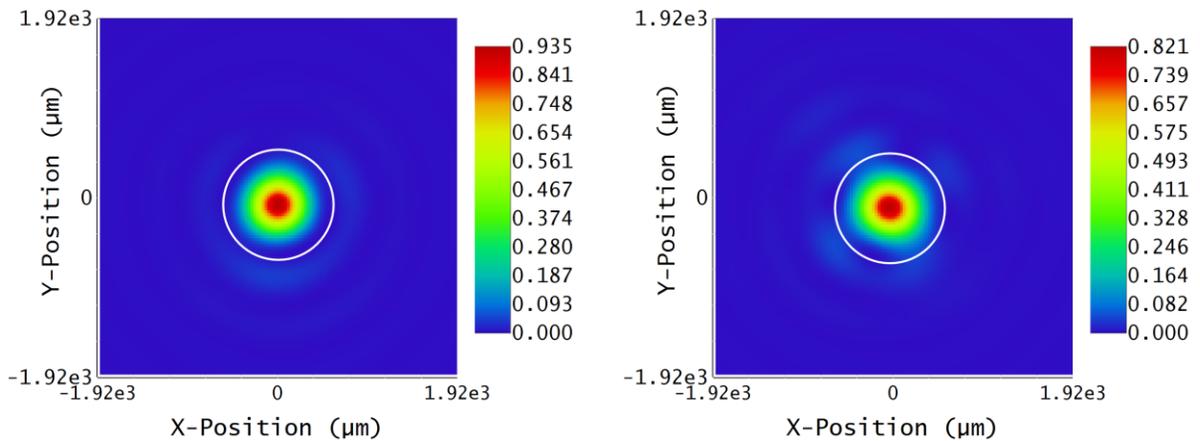

**Fig. 8** The Point Spread Function (PSF) for the best (left) and worst (right) field points of SALTUS demonstrates the system's diffraction-limited performance. The Airy disk (at 30 µm wavelength) is depicted as white circle.

## 3 Tolerancing Analysis

*3.1 Inverse Sensitivity Analysis*

Alignment tolerance analysis was conducted to determine the magnitude of perturbation, as a function of rigid body motion degrees of freedom (DoF), that can be allowed while still meeting the SALTUS optical performance requirements. The results of this tolerancing provide the maximum permissible misalignment error, establishing benchmarks for the telescope's integration and testing process. In the case of SALTUS, the analysis took into account the rigid body motions of each mirror, which include up to six DoF (decentering in three directions and tilting around three axes), or fewer, depending on the mirror's surface shape symmetries as detailed in Table 3. For example, a flat mirror, due to its symmetries, has only three relevant DoFs for perturbation. Additionally, the M6 Field Scanner (a flat mirror) possesses rigid body motion control capabilities, allowing it to adjust its z-location (i.e., translation along the optical axis) to compensate for both its own and the flat M7 mirror's z-directional translation during the commissioning phase. Therefore, the Z-translation of M6 and M7 was excluded from the tolerance table in Table 3. To make this tolerance analysis as conservative as possible, we did not consider the application of the additional compensation mechanisms that are available (e.g., M1 pressure control achieving approximately 0.001 level pressure changes affecting the power of the M1 surface, M3's deformable mirror capability) in this baseline analysis.

The preliminary tolerance values for these parameters were derived using ZEMAX's standard inverse sensitivity analysis procedure, which established baseline values for allowable perturbations. Then, based on the sensitivity information, practically achievable tolerancing ranges for all optics (e.g., ±0.001 inch from a precision CNC machining process and ±0.5 arcsec using electronic autocollimator) were commonly assigned. The definitive tolerance range for each mirror and DoF is presented in Table 3.

Table 3 SALTUS optical tolerancing parameters and values

| Mirror # | Note | DoF | Tolerance Range |
|---|---|---|---|
| **M1** | Inflatable Optic | Decenter x | ±0.025 mm |
| | | Decenter y | ±0.025 mm |
| | | Tilt x | ±0.0001389° |
| | | Tilt y | ±0.0001389° |
| | | Tilt z | ±0.0001389° |
| | | Z translation | ±0.025 mm |
| **M2** | Asphere | Decenter x | ±0.025 mm |
| | | Decenter y | ±0.025 mm |
| | | Tilt x | ±0.0001389° |
| | | Tilt y | ±0.0001389° |
| | | Tilt z | ±0.0001389° |
| | | Z translation | ±0.025 mm |
| **M3** | Low-order DM | Decenter x | ±0.025 mm |
| | | Decenter y | ±0.025 mm |
| | | Tilt x | ±0.0001389° |
| | | Tilt y | ±0.0001389° |
| | | Tilt z | ±0.0001389° |
| | | Z translation | ±0.025 mm |
| **M4** | Folding Flat | Tilt x | ±0.0001389° |
| | | Tilt y | ±0.0001389° |
| | | Z translation | ±0.025 mm |
| **M5** | Asphere | Decenter x | ±0.025 mm |
| | | Decenter y | ±0.025 mm |
| | | Tilt x | ±0.0001389° |
| | | Tilt y | ±0.0001389° |
| | | Tilt z | ±0.0001389° |
| | | Z translation | ±0.025 mm |
| **M6** | Flat Field Scanner (3DoF Compensator) | Tilt x | ±0.0001389° |
| | | Tilt y | ±0.0001389° |
| **M7** | Fast Steering Mirror | Tilt x | ±0.0001389° |
| | | Tilt y | ±0.0001389° |

*3.2 Performance Evaluation using Monte-Carlo Simulation*

Monte-Carlo simulations were performed to statistically confirm the alignment tolerance range, while the final yield and optical performance were monitored. In the simulation process, ZEMAX evaluated 200 random Monte-Carlo misaligned SALTUS configurations with a uniform distribution, based on the parameter ranges detailed in Table 3. The standard deviation for the uniform distribution is defined as σ = (max - min) / (12)$^{1/2}$. The simulation's figure of merit was set as the percentage of Encircled Energy (EE %) within the target radius area.

The wavefront RMS error yield plot in Fig. 9 shows excellent diffraction-limited performance of ~0.07 waves RMS wavefront error with 90% yield at the shortest science wavelength (i.e., 30 µm), which correspond to ~0.81 Strehl Ratio [15]. As additional informative alignment tolerancing figure of merit, -the simulations assessed the average value of the Encircled Energy (EE) within radii of 1 mm, 1.5 mm, and 2 mm at the final focal plane considering various instrument options [14]. Figure 10 shows the yield (i.e., the percentage of configurations meeting the encircled energy figure of merit threshold) as the statistical performance tolerancing simulation result.

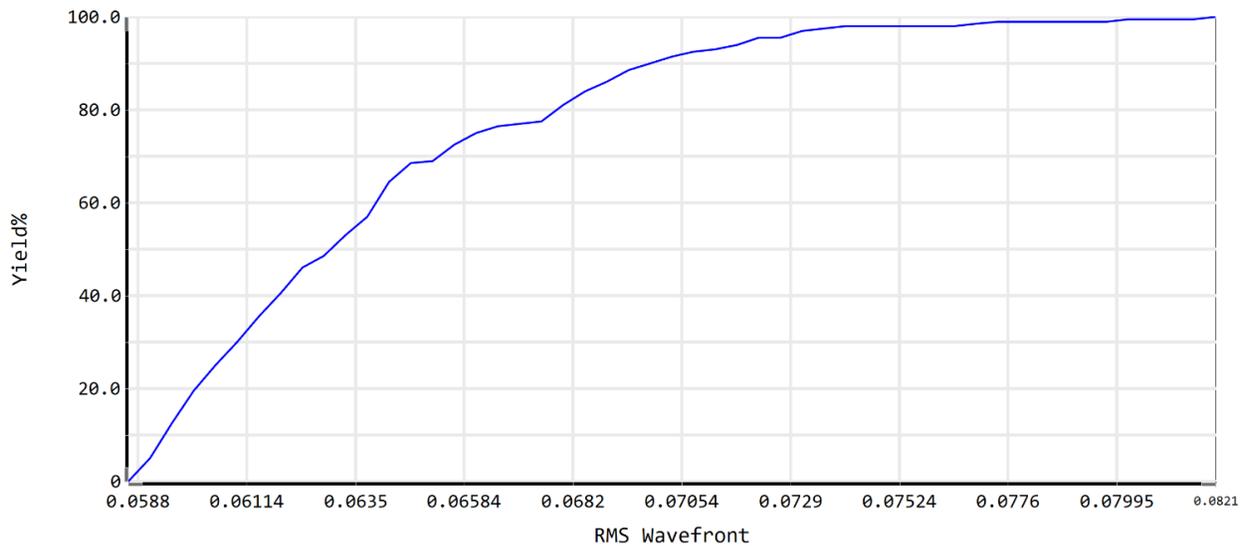

**Fig. 9** The plot shows the yield as a function of the SALTUS total RMS wavefront error (in waves at the 30 µm wavelengths) due to the misalignment tolerance listed in Table 3.

An >90% EE corresponds to ~99% yield on all three yellow, orange, and blue curves, indicating that about 99% of the simulated misaligned cases achieved more than 90% encircled energy within the 1 mm, 1.5 mm, and 2 mm radius at the final focal plane area, meeting the SALTUS instrument requirement. This yield plot offers the SALTUS science mission team trade-off options during the manufacturing and engineering processes, with the EE % goal driven by the requirements of the science instruments.

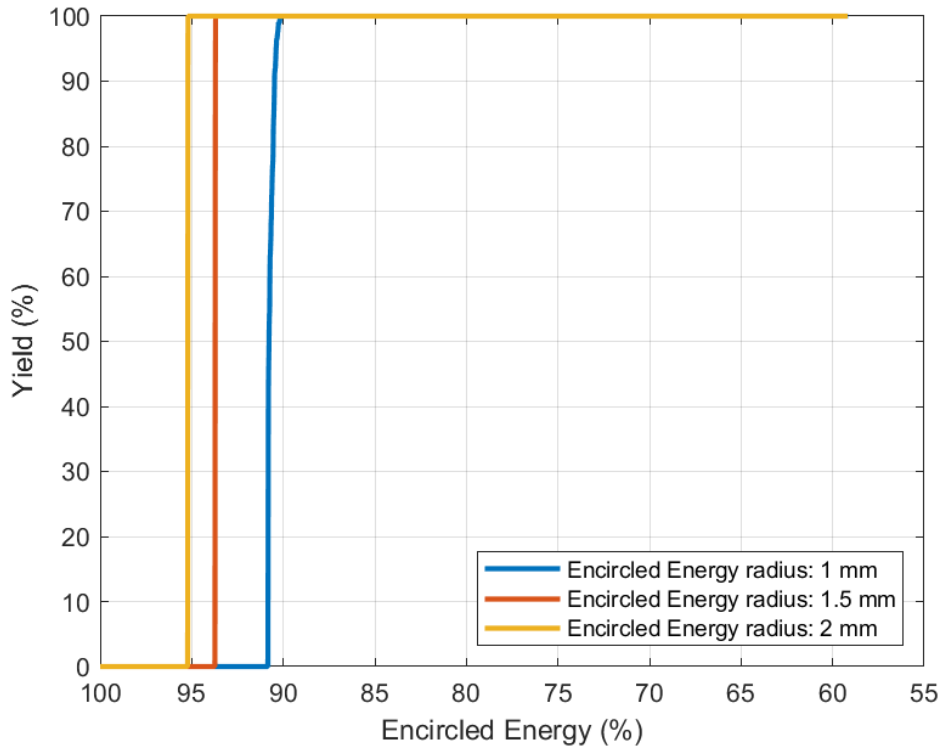

**Fig. 10** The plot shows the yield as a function of the Encircled Energy (EE), serving as the figure of merit, for three different encircled energy radii (1 mm, 1.5 mm, and 2 mm) cases.

Considering SALTUS's customized integration and testing process (as opposed to a mass manufacturing scenario where yield primarily depends on statistical distribution), fine adjustments to balance various aberrations and errors will guarantee an even higher success rate, albeit at the expense of the custom tuning time and cost.

## 4   Non-Sequential Stray Light Analysis

The stray light analysis was performed using non-sequential ray tracing to ensure effective suppression of unwanted light, which could degrade the science signal-to-noise ratio and the observatory's overall operational performance. Stray light primarily originates from surfaces of transmissive elements (i.e., ghost images from Fresnel reflections) or scattering off mirror surface imperfections, such as contamination or micro-roughness.

The ghost light from the inflatable M1's transparent canopy surface was modeled and simulated. Ghost light-1, directly reflected from the membrane canopy surface, diverges, unlike the intended converging path of light reflected from M1, as shown in Fig. 11. Ghost light-2, resulting from multiple reflections between the canopy surface and the M1 surface, is also quickly focused and then diverged (indicated by the red circle region in Fig. 11). A very small fraction of ghost light from the inflatable M1 may arrive at M2, but it does not reach M3 or the subsequent mirrors (M4 to M7), due to the significantly different incidence angles from the nominal rays and does not reach the final focal plane. Therefore, there is no ghost image degradation attributable to Fresnel reflection from the transparent canopy surface of the inflatable structure.

The surface scattering effect, due to micro-roughness and contamination on all seven mirror surfaces (M1 to M7), was also simulated and evaluated using a 1% Lambertian surface scattering model. It is worth noting that the SPDT manufacturing process may create some tool-mark associated surface features (e.g., <~10 nm RMS magnitude), which can be an important factor for visual wavelength (e.g., 0.6 µm) applications requiring <~2 nm RMS micro roughness quality. However, for the >30 µm SALTUS wavelength, the surface quality is sufficiently good as the wavelength is ~50 time longer while the micro-roughness values are only ~5 times larger.

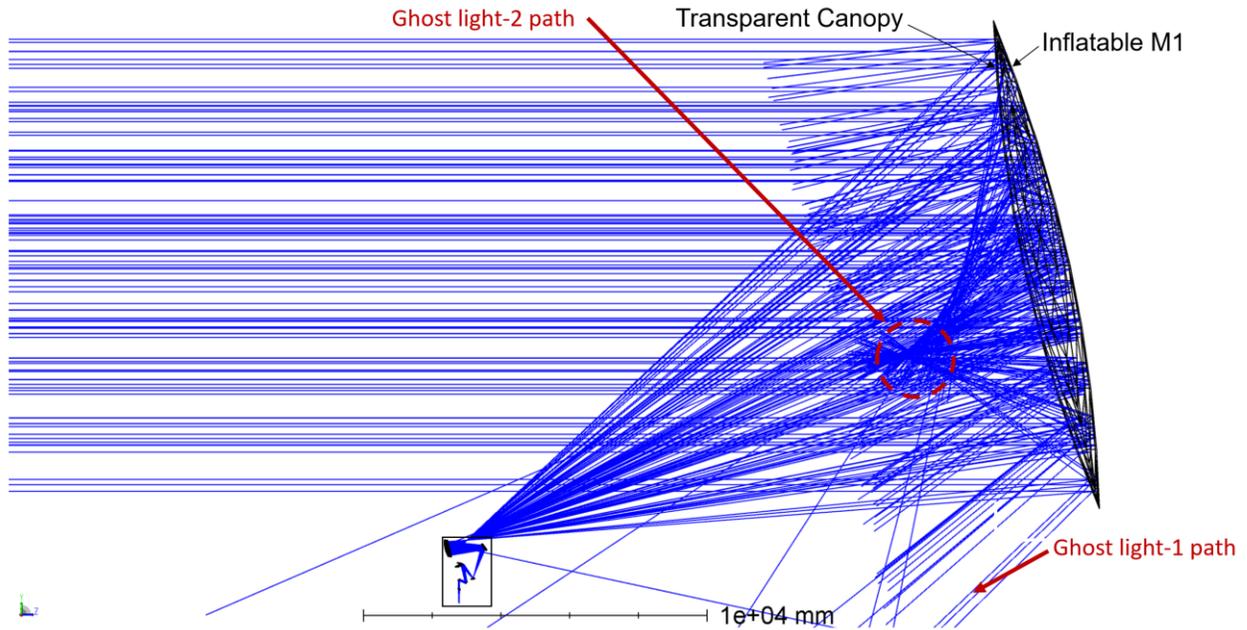

**Fig. 11** Non-sequential optical ray tracing paths were utilized for the SALTUS stray light analysis. Ghost light-1 is directly reflected and diverges from the transparent canopy surface. Ghost light-2 consists of light that reflects multiple times internally between the canopy surface and M1, rapidly converging and then diverging after passing through a near focusing zone (indicated by the red circle).

The surface scattering model predicts a reduction in the irradiance of light entering the science instruments by 3.95%. This conservative (i.e., without any baffles/vanes) background noise performance can be improved by utilizing black baffling and vane design when the opto-mechanical and structural designs are finalized and confirmed. In the meantime, this value serves as a useful reference for making trade-offs between science throughput and overall engineering costs, balancing higher quality mirror manufacturing and contamination control.

## 5 Conclusions

The deployable SALTUS optical design offers an unprecedentedly large photon collecting area at a fraction of the cost and mass budget of traditional space telescope missions. It maintains nominal diffraction-limited performance (i.e., Strehl ratio greater than 0.8) across the entire $\pm\,0.02° \times 0.02°$ Field of View. The adaptable optical system, utilizing the deformable M3 mirror, compensates for on-orbit aberrations as they are deployed, ensuring the quality of far-infrared (far-IR) science observations. Both tolerance and stray light analyses have successfully confirmed the robust and resilient optical performance of the SALTUS design, meeting the requirements of the science instruments. The SALTUS observatory is poised to provide the astrophysics community with a large, unobscured, far-IR space observatory, enabling groundbreaking exploration of our cosmic origins.

*Data Availability*

The data that support the findings of this study are available from the corresponding author upon reasonable request.

*References*

1. Gordon Chin, Carrie Anderson, Jonathan Arenberg (et al), "Single Aperture Large Telescope for Universe Studies (SALTUS): Science Overview," J. Astron. Telesc. Instrum. Syst. (this issue)
2. Leon Harding, Jonathan W Arenberg, Benjamin Donovan (et al), "SALTUS Probe Class Space Mission: Observatory Architecture & Mission Design," J. Astron. Telesc. Instrum. Syst. (this issue)
3. A. B. Meinel and M. P. Meinel, "Inflatable membrane mirrors for optical passband imagery," Opt. Eng., vol. 39, pp. 541–550, 2000, doi: 10.1117/1.602393.
4. W.B. Fichter, "Some Solutions for the Large Deflections of Uniformly Loaded Circular Membranes," NASA Technical Paper 3658, July 1997

5. Arthur L. Palisoc, Gerard Pardoen, Yuzuru Takashima, Aman Chandra, Siddhartha Sirsi, Heejoo Choi, Daewook Kim, Henry Quach, Jonathan W. Arenberg, Christopher Walker, "Analytical and finite element analysis tool for nonlinear membrane antenna modeling for astronomical applications," Proc. SPIE 11820, Astronomical Optics: Design, Manufacture, and Test of Space and Ground Systems III, 118200U (24 August 2021).

6. H. Quach, H. Kang, S. Sirsi, A. Chandra, H. Choi, M. Esparza, K. Karrfalt, J. Berkson, Y. Takashima, et al., "Surface Measurement of a Large Inflatable Reflector in Cryogenic Vacuum," Photonics 9(1), 1, MDPI AG (2021) [doi:10.3390/photonics9010001].

7. Lee D. Feinberg, Bruce H. Dean, William L. Hayden, Joseph M. Howard, Ritva A. Keski-Kuha, Lester M. Cohen, "Space telescope design considerations," Opt. Eng. 51(1) 011006 (6 February 2012) https://doi.org/10.1117/1.OE.51.1.011006

8. Daewook Kim, Heejoo Choi, Ewan Douglas, "Compact three mirror anastigmat space telescope design using 6.5m monolithic primary mirror," Proc. SPIE 12677, Astronomical Optics: Design, Manufacture, and Test of Space and Ground Systems IV, 126770E (4 October 2023); https://doi.org/10.1117/12.2682180

9. W. D. Henderson, S. V. Gunn, "System Performance Of A Large Deformable Mirror Using Differential Ball Screw Actuators," Proc. SPIE 0179, Adaptive Optical Components II, (11 July 1979); doi: 10.1117/12.957293

10. Vorontsov, Mikhail A., and Viktor P. Sivokon. "Stochastic parallel-gradient-descent technique for high-resolution wave-front phase-distortion correction." JOSA A 15, no. 10 (1998): 2745-2758.

11. Chen, Erhu, Hongwei Cheng, Yinbing An, and Xiaofeng Li. "The improvement of spgd algorithm convergence in satellite-to-ground laser communication links." Procedia Engineering 29 (2012): 409-414.

12. Lei Huang, Heejoo Choi, Wenchuan Zhao, Logan R. Graves, and Dae Wook Kim, "Adaptive interferometric null testing for unknown freeform optics metrology," Opt. Lett. 41, 5539-5542 (2016).

13. Lei Huang, Chenlu Zhou, Wenchuan Zhao, Heejoo Choi, Logan Graves, Dae Wook Kim, "Close-loop performance of a high precision deflectometry controlled deformable mirror (DCDM) unit for wavefront correction in adaptive optics system," Optics Communications 393, 83-88 (2017).

14. J. R. G. Silva, B. Mirzaei, W. Laauwen, N. More, A. Young, C. Kulesa, C. Walker, A. Khalatpour, Q. Hu, C. Groppi, J. R. Gao, "4×2 HEB receiver at 4.7 THz for GUSTO," Proc. SPIE 10708, Millimeter, Submillimeter, and Far-Infrared Detectors and Instrumentation for Astronomy IX, 107080Z (9 July 2018); https://doi.org/10.1117/12.2313410

15. Virendra N. Mahajan, "Strehl ratio for primary aberrations: some analytical results for circular and annular pupils," J. Opt. Soc. Am. 72, 1258-1266 (1982)